\documentclass[12pt,preprint]{aastex}

\begin{document}

\shorttitle{Black Holes}
\shortauthors{Narayan}

\title{Black Holes in Astrophysics}

\author{Ramesh Narayan} \affil{Harvard College Observatory,
Harvard-Smithsonian Center for Astrophysics, Cambridge, MA 02138, USA}
\email{rnarayan@cfa.harvard.edu}

\begin{abstract}
This article reviews the current status of black hole astrophysics,
focusing on topics of interest to a physics audience.  Astronomers
have discovered dozens of compact objects with masses greater than
$3M_\odot$, the likely maximum mass of a neutron star.  These objects
are identified as black hole candidates.  Some of the candidates have
masses $\sim 5 - 20M_\odot$ and are found in X-ray binaries, while the
rest have masses $\sim10^6-10^{9.5}M_\odot$ and are found in galactic
nuclei.  A variety of methods are being tried to estimate the spin
parameters of the candidate black holes.  There is strong
circumstantial evidence that many of the objects have event horizons,
so there is good reason to believe that the candidates are true black
holes.  Recent MHD simulations of magnetized plasma accreting on
rotating black holes seem to hint that relativistic jets may be
produced by a magnetic analog of the Penrose process.

\end{abstract}


\section{Introduction}

While physicists have been grappling with the theory of black holes
(see the many articles in this volume), astronomers have been
searching for real-life examples of black holes in the universe.  This
search has been remarkably successful.  Dozens of excellent black hole
candidates have been identified, and their properties are being
investigated with a variety of techniques.

Black hole (BH) astrophysics began in the 1960s with the discovery of
quasars and other active galactic nuclei (AGN) in distant galaxies.
Early on it became clear that the most natural explanation for the
extraordinary luminosity of AGN is the release of gravitational energy
through accretion of gas onto supermassive BHs (see Rees 1984 for a
review).  BH astrophysics received a boost in the 1970s with the
discovery of the X-ray binary (XRB) Cygnus X--1 (Bolton 1972).  The
X-ray-emitting compact star in this binary seemed to be unusually
massive, so it was inferred to be a BH (e.g., Gies \& Bolton 1986).
The next important breakthrough came with the work of McClintock \&
Remillard (1986), who showed that the source A0620--00, a member of a
class of XRBs known as X-ray novae, has a mass almost certainly
greater than the maximum mass of a neutron star (NS).  This important
discovery opened the way to many additional BH candidates being found
in other X-ray novae.  Today about 20 excellent BH candidates are
known in XRBs (see McClintock \& Remillard 2004 for a recent review).
In the last decade, supermassive BHs have returned to the fore as
astrophysicists have obtained impressive dynamical evidence for the
presence of compact, massive, dark objects in the nuclei of nearby
galaxies.

\section{Basic Concepts}

Astrophysical BHs are macroscopic objects with masses ranging from
several $M_\odot$ (in XRBs) to $10^6-10^{9.5}M_\odot$ (in galactic
nuclei), where $1M_\odot = 1.99\times10^{30} ~{\rm kg} = $ the mass of
the Sun.  Being so massive, these BHs are described completely by
classical general relativity.  As such, each BH is characterized by
just three numbers: mass $M$, spin parameter $a_*$, defined such that
the angular momentum of the BH is $a_*GM^2/c$, and electric charge $Q$
(Misner, Thorne \& Wheeler 1973; Shapiro \& Teukolsky 1983; Hartle
2003).  Actually, an astrophysical BH is not likely to have any
significant electric charge because it will usually be rapidly
neutralized by surrounding plasma.  Therefore, the BH can be fully
characterized by measuring just two parameters, $M$ and $a_*$, of
which the latter is constrained to lie in the range 0 (non-spinning
BH) to 1 (maximally-spinning BH).

Using the known equation of state of matter up to nuclear density and
applying general relativity, it can be shown that the maximum mass of
a compact relativistic star such as a NS is $\sim2-3M_\odot$
(Nauenberg \& Chapline 1973; Sabbadini \& Hartle 1973; Rhoades \&
Ruffini 1974).  This result allows the following simple and powerful
technique for identifying BH candidates:

\noindent
{\it Find a compact (relativistic) astrophysical object and measure
its mass.  If the mass is greater than about $3M_\odot$, then the
object is very likely a BH.}  

\noindent
\S 3 reviews the candidate BHs that have been discovered so far by
this technique.  What is remarkable is that in the process of
identifying a BH candidate one has already measured $M$, one of the
two parameters characterizing the object.  All that is left then is to
measure $a_*$ (easier said than done, as we discuss in \S 4) and we
will know everything there is to know about the object!

General relativity makes very precise predictions about the properties
of space-time exterior to the horizon of a stationary BH.  For the
general case of a spinning (uncharged) BH, the space-time is described
by the Kerr metric, which simplifies to the Schwarzschild metric when
$a_*=0$ (Misner et al. 1973; Hartle 2003).  Given the metric, one can
calculate the orbits of test particles around the BH.  These could, in
principle, be used to interpret observations and to estimate the spin
of the BH (\S4).

Of all the extraordinary characteristics of a BH none is more
remarkable than the fact that it possesses an {\it event horizon} ---
a one-way membrane that causally isolates the ``inside'' of the BH
from the rest of the universe.  Matter, radiation, energy,
information, etc. can fall into the BH from the outside, but nothing,
not even light, can get out (at least within classical physics).  An
important question is whether the BH candidates identifed by
astronomers conform to this remarkable prediction of general
relativity:

\noindent
{\it Do astrophysical BH candidates possess event horizons?}

\noindent
Considerable progress has been made towards answering this question,
as discussed in \S 5.  

The radius of the event horizon of a non-spinning BH is given by the
Schwarzschild radius,
\begin{equation}
R_S = {2GM\over c^2} = 2.95\,\left({M\over M_\odot}\right) ~{\rm km}.
\end{equation}  
The radius is smaller in the case of a spinning BH, tending to
$GM/c^2$ as $a_* \to 1$.

A most interesting property of a spinning BH is that it has an {\it
ergosphere} (Misner et al. 1973; Hartle 2003).  This is a region
external to the horizon where the dragging of inertial frames by the
spinning hole is so strong that the only allowed orbits are those that
corotate with the BH (as seen from an intertial frame at infinity).
Penrose (1969) showed that it is possible to set up particle orbits
inside the ergosphere of a spinning BH so as to extract rotational
energy from the BH.  Does the Penrose process have any application in
astrophysics?  This question is discussed in \S6.

\section{Measuring Black Hole Mass}

The most accurate mass measurements in astrophysics are made via
dynamical methods.  Consider a test particle that is in a circular
orbit around a mass $M$.  If the orbit is wide enough for Newtonian
physics to apply, then we have
\begin{equation}
M = {v^2r\over G} = {4\pi^2r^3\over GP_{\rm orb}^2} = {v^3P_{\rm
orb}\over 2\pi G},
\end{equation}
where $r$ and $P_{\rm orb}$ are the radius and period of the orbit and
$v$ is the velocity of the particle.  By measuring any two of $v$, $r$
and $P_{\rm orb}$, we may estimate $M$.  Equation (2) is modified in a
straightforward way when the orbit is non-circular.  For instance, the
relation $M=4\pi^2r^3/GP_{\rm orb}^2$ continues to be valid, provided
$r$ is taken to be the semi-major axis of the elliptical orbit of the
test particle.

\subsection{X-ray Binaries}

In the case of a BH XRB, it is relatively easy to measure the period
$P_{\rm orb}$ of the orbit and the maximum line-of-sight Doppler
velocity $K_c = v\sin i$ of the companion star to the BH, where $i$ is
the inclination angle of the binary orbit.  From these, one can
calculate the ``mass function'' $f(M)$:
\begin{equation}
f(M) \equiv {K_c^3P_{\rm orb}\over 2\pi G} = {M \sin^3i \over
(1+M_c/M)^2},
\end{equation}
where $M$ is the mass of the BH candidate and $M_c$ is the mass of the
companion star.  Comparing equation (3) with equation (2) we see two
differences.  First, because we measure only the line-of-sight
component of the orbital velocity, the mass function differs from the
BH mass by a factor of $\sin^3 i$.  Often, we have an independent
estimate of $i$ from the the light curve of the binary, but this
estimate is rarely as reliable as the measurements of $K_c$ and
$P_{\rm orb}$.  The second difference is that the secondary is not a
test particle but has a finite mass.  This gives the factor of
$(1+M_c/M)^2$ in the denominator, which corrects for the reflex motion
of the BH.  Therefore, one requires an independent estimate of $M_c$
before one can calculate $M$.  In XRBs like Cyg X--1, $M_c$ is large
and seriously affects our ability to estimate $M$.  In contrast,
systems like A0620--00 and other X-ray novae have low-mass companions
which often have masses less than 10 percent of $M$.  In these
systems, even a very approximate estimate of $M_c$ is sufficient.

By combining measurements of $K_c$ and $P_{\rm orb}$ with estimates of
$i$ and $M_c$, the masses of the compact X-ray-emitting stars in a
number of XRBs have been measured.  Presently, 20 XRBs have been shown
to have $M>3M_\odot$ and these objects are all excellent BH candidates
(McClintock \& Remillard 2004).  However, the mass estimates are
reliable only if $i$ (and to a lesser extent $M_c$) is well
determined, which is not always the case.  Fortunately, an inspection
of equation (3) shows that the mass function $f(M)$, which depends
only on the two accurately measured quantities $K_c$ and $P_{\rm
orb}$, is a strict lower bound on $M$.  Most of the 20 XRBs have
$f(M)$ itself greater than or of order $3M_\odot$.  Therefore, these
systems are excellent BH candidates, regardless of uncertainties in
their inclinations and companion star masses.

\subsection{Galactic Nuclei}

In the center of our Milky Way Galaxy is a dark massive object whose
existence has been inferred for a number of years from its effect on
the motions of stars and gas in its vicinity.  Recently, high
resolution infrared observations have enabled two independent groups
to follow the orbits of individual stars around this object (Sch\"odel
et al. 2002; Eisenhauer et al. 2003; Ghez et al. 2003, 2004).  Movie1
and Movie2 show time-elapsed images of the Galactic Center region
revealing the (eccentric) orbits of several stars.  By modeling the
orbits with Newtonian dynamics --- essentially using the
generalization of equation (2) for elliptical orbits --- the mass of
the dark object has been estimated to be $3.7\pm 0.2\times10^6
M_\odot$.  The mass must lie within about $2\times10^{13}$ m (the
distance of closest approach of the observed stars), strongly
suggesting that the object is a BH.  If the object is not a BH, then
the only reasonable hypothesis is that it is a dense cluster of dark
compact stars.  However, such a cluster would be short-lived and would
become a BH in much less time than the age of the Galaxy (Maoz 1998),
so this is not a very likely scenario.

A famous radio source called Sagittarius A* (Sgr A*) is located very
close to the position of the dark mass described above; it has for
long been suspected to be the supermassive BH.  Reid \& Brunthaler
(2004) have followed the motion of Sgr A* over a period of eight years
using radio interferometry and have shown that the component of its
velocity perpendicular to the plane of the Galaxy is $-0.4\pm0.9 ~{\rm
km\,s^{-1}}$, consistent with zero.  Since Sgr A* should experience
Brownian motion as a result of gravitational interactions with stars
in its vicinity, its lack of apparent motion implies that it has a
large mass.  Depending on the statistical method that one employs and
the confidence level that one seeks, one infers that Sgr A* must have
a mass of at least $10^5 M_\odot$; in fact, Sgr A* most likely
encompasses the entire dark mass in the Galactic Center.  The image of
Sgr A* as measured in millimeter radio waves indicates an angular size
of about 240 micro-arcsecond (Bower et al. 2004), which corresponds to
$27 R_S$ for a BH mass of $3.7\times10^6M_\odot$ and a distance of 8
kpc.  Thus, Sgr A* has both a large mass and a very small radius.
Surely it must be a BH!

The nearby galaxy NGC 4258 is another remarkable object, with a disk
of gas in its nucleus that emits radio waves via maser emission from
water molecules (Miyoshi et al. 1995; Greenhill et al. 1995;
Herrnstein et al. 1998).  Radio interferometry measurements have shown
that the gas follows circular orbits with a nearly perfect Keplerian
velocity profile ($v \propto r^{-1/2}$, see Fig. 1).  Furthermore, the
acceleration of the gas has been measured and it too is consistent
with Keplerian dynamics (Bragg et al. 2000).  From these measurements
it is inferred that there is a dark object with a mass of
$3.5\times10^7 M_\odot$ confined within $\sim4\times10^{15}$ m of the
center of NGC 4258.  The case for this dark mass being a BH is again
extremely strong.

Supermassive BHs have been inferred in the nuclei of many other nearby
galaxies by means of optical observations, especially with the Hubble
Space Telescope (Kormendy \& Richstone 1995; Pinkney et al. 2003).
One or two galaxies have coherent gas disks at the center whose
orbital velocities provide information on the enclosed mass.  For the
rest, one measures the velocity dispersion of stars near the center of
the galaxy and applies the virial theorem --- essentially a
statistical version of equation (2) --- to infer the enclosed mass.
Masses in the range $10^6M_\odot$ to $3\times10^{9.5}M_\odot$ have
been estimated by this means in about 20 galaxies.  Although the
constraints on the radii of the dark mass concentrations in these
galactic nuclei is relatively poor (compared to Sgr A* or NGC 4258),
nevertheless the case for identifying the objects as supermassive BHs
is strong.

Central black hole masses are difficult to measure for more distant
galaxies.  Techniques such as ``reverberation mapping'' (see Horne et
al. 2004) --- another application of the virial theorem --- can be
used if the black hole is an AGN and shows variability.  Such methods
are less accurate than the ones described above, but they are valuable
since they provide black hole mass estimates for a large sample of
galaxies.

\subsection{The Mass Distribution of Astrophysical Black Holes}

The two categories of BHs described above are clearly very distinct
from each other, with very different masses: $M \sim {\rm few}
-20M_\odot$ for stellar-mass BHs in XRBs and $M\sim
10^6-10^{9.5}M_\odot$ for supermassive BHs in AGN.  The BHs in XRBs
are clearly the remnants of very massive stars (say with initial
masses $M>30M_\odot$) at the end of their lives.  But how exactly are
the BHs in galactic nuclei formed, and how do they evolve?  

Observations have revealed interesting correlations between the
supermassive BHs in the nuclei of galaxies and the properties of their
host galaxies.  For instance, the BH mass appears to be roughly 0.1\%
of the mass of the galaxy bulge (Magorrian et al. 1998; H\"aring \&
Rix 2004), and there is an even tighter correlation between the BH
mass and the velocity dispersion of the bulge (Ferrarese \& Merritt
2000; Tremaine et al. 2002); the latter quantity is, of course, a
measure of the galaxy mass via the virial theorem.  These correlations
suggest that there is an intimate connection between the supermassive
BH and the galaxy, even though the BH constitutes only a small
fraction of the galaxy in terms of mass.  A number of explanations
have been offered for the correlations (e.g., Rees \& Silk 1998;
Murray, Quataert \& Thompson 2005), but there is no real understanding
of the correlation at present.

Are the two kinds of BHs discussed above the only ones in the universe
or are there other kinds, e.g., intermediate mass BHs with masses of
say $10^3-10^4M_\odot$?  This question has attracted recent attention.
In several nearby galaxies there is a class of ultra-luminous X-ray
sources (Fabbiano 1989; Colbert \& Mushotzky 1999) which seem to be
too bright to be ordinary $10M_\odot$ BHs; some of these sources have
luminosities of $10^{41} ~{\rm erg\,s^{-1}}$ or more, whereas the
nominal maximum steady luminosity of a gravitating object is the
Eddington limit, $L_{\rm Edd} =1.3\times 10^{39}(M/10M_\odot) ~{\rm
erg\,s^{-1}}$.  The Eddington limit is the luminosity at which the
outward acceleration of gas by radiation pressure is just equal to the
inward acceleration by gravity.  For luminosities greater than $L_{\rm
Edd}$, the radiation force is expected to overwhelm gravity and to
cause the accretion rate to reduce such that the luminosity falls
below $L_{\rm Edd}$ (see Shapiro \& Teukolsky 1983 for additional
discussion).  Although there are ways around the Eddington limit,
these require special conditions; so there is a good chance that at
least some of the ultraluminous X-ray sources are much more massive
than $10M_\odot$.  On the other hand, the sources are not likely to be
(underluminous) $10^6M_\odot$ BHs because they are generally located
away from the nuclei of their host galaxies.  Also, in some galaxies,
several ultra-luminous sources have been found whereas only one
supermassive BH is typically found in a galaxy.

It is unclear at present what exactly the ultra-luminous X-ray sources
are and whether they are even a single homogeneous population (see
Miller \& Colbert 2004 for a review).  Dynamical mass measurements
would obviously settle the issue.  Unfortunately, none of the sources
has a confirmed binary companion, so there is no prospect of making a
robust mass measurement any time soon.  If ultraluminous X-ray sources
are ultimately confirmed to be intermediate mass BHs, an interesting
question would then need to be settled: are they a distinct new
population or are they just an extended tail of the mass distribution
of either stellar-mass BHs or supermassive BHs?

\section{Estimating Black Hole Spin}

Estimating the mass of a BH is relatively easy since mass has a
measurable effect even at large radii, where Newtonian gravity
applies.  Spin, on the other hand, does not have any Newtonian effect
--- the orbit of a planet in the solar system, for instance, would be
the same whether the Sun corotates or counter-rotates with the
planetary orbit.  Only for relativistic orbits does spin have
measurable effects.  Therefore, to measure $a_*$, we need test
particles on orbits with very small radii.  Fortunately, astrophysical
BHs do have such test particles in the form of accreting gas.

When considering circular orbits in a BH spacetime (e.g., Misner et
al. 1973; Shapiro \& Teukolsky 1983; Hartle 2003), a key concept is
the innermost stable circular orbit, or ISCO, with a radius designated
$R_{\rm ISCO}$ (also referred to as the marginally stable orbit).
Circular orbits with radii $R \geq R_{\rm ISCO}$ are stable to small
perturbations, whereas those with $R < R_{\rm ISCO}$ are unstable.
Figure 2 shows the variation of $R_{\rm ISCO}$ with $a_*$.  For a
maximally spinning BH, $R_{\rm ISCO} = GM/c^2$ if the orbit corotates
with the BH ($a_*=+1$ in Fig. 2) and $R_{\rm ISCO}=9GM/c^2$ if it
counter-rotates ($a_*=-1$); for a non-spinning BH ($a_*=0$), $R_{\rm
ISCO}=6GM/c^2$.  Corresponding to changes in $R_{\rm ISCO}$, there are
variations in the angular velocity of an orbiting particle at $R_{\rm
ISCO}$ (as measured at infinity), and in the binding energy of the
particle.  These variations are shown in Figure 2.

The gas in an accretion disk starts from large radii and spirals in
through a sequence of nearly circular orbits as it viscously loses
angular momentum.  When the gas reaches the ISCO, no more stable
circular orbits are available, so the gas accelerates radially and
free-falls into the BH.  Thus, the ISCO serves effectively as the
inner edge of the accretion disk.  A variety of observational methods
have been proposed for estimating the radius $R_{\rm in}=R_{\rm ISCO}$
of the disk inner edge, or one of the other quantities plotted in
Figure 2, with a view to thereby estimating $a_*$.

\subsection{Spectral Fitting}

When a BH has a large mass accretion rate $\dot M$, corresponding to
an accretion luminosity $L_{\rm acc}$ above a few per cent of $L_{\rm
Edd}$, the accreting gas tends to be optically thick and to radiate
approximately as a blackbody.  In this spectral state, called the
``high soft state'' (McClintock \& Remillard 2004), one can
theoretically calculate the flux of radiation $F(R)$ emitted by the
accretion disk, and hence obtain the effective temperature profile
$T_{\rm eff}(R) \equiv [F(R)/\sigma]^{1/4}$, where $\sigma$ is the
Stefan-Boltzmann constant.  If the disk emits as a true blackbody at
each radius, it is a simple matter to calculate the total spectral
luminosity $L_\nu d\nu$.  By comparing this quantity with the spectral
flux $F_\nu d\nu$ received at Earth, one obtains an estimate of
$R_{\rm in}^2\cos i/D^2$ (essentially the projected solid angle of the
disk), where $i$ is the inclination angle and $D$ is the distance to
the source.

In a few BH XRBs, sufficiently reliable estimates of $i$, $D$ and $M$
are available to carry out this exercise and thus to estimate $R_{\rm
in}/(GM/c^2)$.  If the inner edge of the disk is at the ISCO, as is
very likely in the ``high state,'' one then obtains $a_*$ (via
Fig. 2).  This method was first used by Zhang, Cui \& Chen (1997, see
also Sobczak et al. 1999) for the XRBs GRO J1655--40 and GRS 1915+105,
and more recently by Li et al. (2004) for the source 4U1543--47.

A major weakness of this method is that a number of effects ---
frequency-dependent opacity, Thomson/Compton scattering, etc. --- will
cause the spectrum of an accretion disk to deviate from a blackbody.
It is traditional to model these effects by retaining the blackbody
assumption but artificially increasing the temperature of the emitted
radiation by a spectral hardening factor $f\sim1.7$ (Shimura \&
Takahara 1995).  The problem is that the correction is very
approximate.  This might still be tolerable if the magnitude of the
correction were small, but in fact the correction is quite large.
Therefore, spin estimates obtained by this method should be treated
with caution.  Also, the method requires accurate estimates of $M$,
$i$ and $D$, whereas the next two methods described below require
fewer parameters to be measured.

\subsection{Quasi-Periodic Oscillations}

For several BH XRBs, the power spectrum of intensity variations shows
one or two peaks (more like bumps in some cases) at frequencies of a
few hundred Hz.  The peaks are relatively broad, indicating that they
do not correspond to coherent oscillations but rather to
quasi-periodic oscillations (QPOs) with a fairly low quality factor
$Q$ (McClintock \& Remillard 2004).  The observed high frequency
suggests that the oscillations arise in gas that is close to the BH.
This gas is presumably strongly influenced by relativistic effects,
including in particular effects associated with the spin of the BH.
However, there is at present no clear understanding of exactly what
causes the oscillations.  They could be trapped vibration modes of the
disk (Okazaki, Kato \& Fukue 1987; Nowak \& Wagoner 1991) or
characteristic periods of test particle orbits (Stella \& Vietri 1998)
or vibrations at the interface between gas and magnetic fields (Li \&
Narayan 2004), or something else altogether.  An important clue is
that frequencies in the ratio of 3:2 are often seen, so it would
appear that there is a resonance of some kind between different modes
(Abramowicz \& Kluzniak 2001).

One possibility that has attracted attention is that the QPO with the
highest frequency in each BH XRB corresponds to the orbital
(Keplerian) frequency of gas blobs at some characteristic radius; it
is plausible that this radius corresponds to the inner edge of the
disk.  Since the Keplerian frequency is proportional to $1/M$ and
moreover depends on $a_*$ (assuming that $R_{\rm in} =R_{\rm ISCO}$,
Fig. 2), one can use this method to estimate $a_*$ provided an
estimate of $M$ is available.  The method has been applied to a few BH
XRBs (e.g., Strohmayer 2001).  Recently, there has been tentative
evidence for QPOs with a period of 17 minutes in the infrared emission
from Sgr A*, the supermassive BH in the Galactic Center (Genzel et
al. 2003).  If the QPOs correspond to the Keplerian frequency at any
radius $R>R_{\rm ISCO}$, then the BH must be spinning with $a_*>0.5$.

There is at present no clear evidence that the observed QPOs do
correspond to the Keplerian frequency (for instance, they may be
related to a precession frequency, see Cui, Zhang \& Chen 1998; Stella
\& Vietri 1998).  Nor is there any guarantee that the gas producing
the variations is located at $R_{\rm ISCO}$.  Both assumptions are
certainly plausible, but it would be nice to have corroborating
evidence.  On the other hand, a strength of this method is that it
does not require estimates of $i$ or $D$ --- all that is needed is an
accurate estimate of $M$ and a physical model of the oscillations.  As
described in \S3, mass estimates are generally quite reliable, so the
future of this method depends very much on establishing a quantitative
physical model of the QPOs.

\subsection{Relativistic Iron line}

Tanaka et al. (1995) discovered a strong broad spectral line in the
X-ray spectrum of the AGN MCG--6--30--15.  They interpreted the line
as fluorescent iron K$\alpha$ emission from cool gas in the accretion
disk.  Similar broad lines have been seen in a few other AGN and XRBs
(see Reynolds \& Nowak 2003 for a review).  Whereas the rest energy of
the iron line is 6.4 keV, the observed line extends from about $4-7$
keV, the result of broadening by Doppler blue- and red-shifts as well
as gravitational redshift.

The line width --- indeed the entire shape of the line --- depends on
several factors: (i) The radius range over which the emission occurs
is clearly relevant, especially the position of the innermost radius
of the disk; since $R_{\rm in} = R_{\rm ISCO}$ depends on $a_*$, the
BH spin has an important effect.  (ii) The disk inclination $i$ is
also important, since Doppler effects are much larger for an edge-on
disk than for a face-on disk.  (iii) The line emissivity as a function
of $R$ strongly influences the line shape; the emissivity depends on
the pattern of hard X-ray irradiation of the disk surface for which
there is no reliable theory available at present.

Given a system with a broad iron line, and assuming that the radiating
gas follows Keplerian orbits with radii $R \geq R_{\rm ISCO}$, one can
fit the shape of the line profile by adjusting $a_*$, $i$ and the
emisivity function; the latter is usually modeled as a power-law in
radius, $R^{-\beta}$.  Reynolds \& Nowak (2003) show how the line
shape varies with $a_*$, $i$ and $\beta$.  With good signal-to-noise
data, it is clear that one would be able to fit all three parameters.
The effect of $a_*$ is particularly dramatic.  As the BH spin
increases, the inner edge of the disk comes closer to the horizon
(Fig. 2) and the velocity of the gas increases substantially.  This
gives a wider range of Doppler shifts, as well as a larger
gravitational redshift.  As a result the line extends down to very low
energies below 4 keV (especially when $\beta$ is large).  The
detection of such extreme levels of broadening may be taken as strong
indication of a rapidly spinning BH.

This method of estimating the BH spin has some weaknesses.  First,
there is no guarantee that the emissivity varies with radius as a
simple power-law.  Second, the method depends crucially on the
assumption that the line emission cuts off abruptly inside $R_{\rm
ISCO}$.  However, the gas in the plunging region inside $R_{\rm ISCO}$
may also produce fluorescent emission (since it is likely to be
irradiated just as much as the gas outside $R_{\rm ISCO}$).  Thus one
could have highly redshifted emission even without a rapidly spinning
BH (Reynolds \& Begelman 1998).  On the other hand, the method does
not require any knowledge of the BH mass or the distance, and it
solves for the disk inclination $i$ using the same line data from
which $a_*$ is estimated.  These are important advantages.

In the case of MCG--6--30--15, different observations have revealed
different shapes for the line, suggesting that the line is produced by
a highly variable mechanism.  The data confirm that the emission comes
from a relativistic disk (e.g., Vaughan \& Fabian 2004) and at least
some of the datasets can be interpreted in terms of a rapidly spinning
BH, e.g., assuming that there is no emission from within the ISCO,
Reynolds et al. (2004b) estimate $a_*>0.93$.  Among BH XRBs, the
source GX 339-4 shows a broad iron line which seems to indicate
$a_*>0.8$ (Miller et al.  2004).  Dovciak, Karas \& Yaqoob (2004)
discuss some caveats in using the iron line to estimate $a_*$.

Looking to the future, given a sufficiently clean system and an X-ray
telescope with high sensitivity (e.g., the proposed Constellation-X
and XEUS satellite missions), one might imagine not just measuring
$a_*$ of a BH but even detecting higher-order effects and thereby
testing general relativity.  It is not known how common broad iron
lines are in the X-ray spectra of AGN and XRBs --- some well-studied
sources show clear narrow iron lines and a puzzling absence of any
broad line (Pounds et al. 2003; Reynolds et al. 2004a) --- and
therefore it is unclear how many favorable sources will be available
for serious modeling.  Also, the variability of the line with time
means that it will be challenging to make fundamental tests of gravity
with this method.  On the other hand, the variability could provide
interesting opportunities to study disk dynamics and turbulence, as
Movie3 and Movie4 show.

\subsection{Radiative Efficiency of AGN}

The radiative efficiency of an accretion disk is defined as the energy
it radiates per unit accreted mass: $\eta = L_{\rm acc}/ \dot Mc^2$.
It is determined by the binding energy of gas at the ISCO, which
depends on $a_*$ (Fig. 2).  For a non-rotating BH, $\eta = 0.057$,
whereas a corotating disk around a maximally-rotating BH ($a_*=1$) has
a much higher efficiency, $\eta=0.42$.

In a typical accretion system, one can easily measure $L_{\rm acc}$
(provided the distance is known), but one practically never has an
accurate estimate of $\dot M$, so one cannot calculate $\eta$ with the
precision needed to estimate $a_*$.  However, there is one situation
in which it is possible.  From observations of high redshift AGN, one
can estimate the mean energy radiated by supermassive BHs per unit
volume of the universe.  Similary, by taking a census of supermassive
BHs in nearby galaxies, one can estimate the mean mass in BHs per unit
volume of the current universe.  Assuming that supermassive BHs
acquire most of their mass via accretion (a not unreasonable
hypothesis), one can divide the two quantities to obtain the mean
radiative efficiency of AGN.  This is a variant of an idea originally
due to Soltan (1982; see also Merritt \& Ferrarese 2001).  The current
data suggest an efficiency $\eta\sim0.1-0.15$ for supermassive BHs on
average (Elvis, Risaliti \& Zamorani 2002; Yu \& Tremaine 2002), and
$\sim0.2$ or even larger for the most massive systems (Yu \& Tremaine
2002).  Such large values of $\eta$ are possible only if supermassive
BHs have significant rotation.

There are several uncertainties in the above argument, but many of
them only cause $\eta$ to be underestimated.  Therefore, it does
appear that supermassive BHs as a class must have substantial
rotation.  However, it should be noted that this is only a statistical
result for the population as a whole, and the method does not say
anything about the rotation of any specific BH.

\section{Evidence for the Event Horizon}

While the measurements of mass and spin described in the previous
sections are very impressive, it is important to keep in mind that
there is as yet no definite proof that any of the BH candidates
discovered so far is truly a BH.  All we know is that the objects are
too massive ($>3M_\odot$) to be NSs.  But mass measurements by
themselves cannot prove that the objects are BHs.

To prove that an object is a BH requires an unambiguous demonstration
that it possesses an event horizon, a challenging (perhaps technically
impossible, Abramowicz, Kluzniak \& Lasota 2002) undertaking.
Nevertheless, considerable progress has been made, and there is strong
circumstantial evidence that many of the BH candidates discovered so
far do possess event horizons.

\subsection{Luminosities of Quiescent BHs}

When a BH accretes at a rate much less than the Eddington rate, the
radiative efficiency of the accreting gas is found to be very small,
i.e., the efficiency parameter $\eta=L_{\rm acc}/\dot Mc^2$ defined
earlier is found to be $\ll1$.  The best-known example is Sgr A*,
which is extremely underluminous, $L_{\rm acc} \sim 10^{36} ~{\rm
erg\,s^{-1}}$, compared to the Eddington luminosity $L_{\rm Edd} \sim
5\times10^{44} ~{\rm erg\,s^{-1}}$ for its mass.  What is interesting
is that there is considerable gas available around Sgr A*, and if the
accretion flow possessed a standard efficiency $\eta\sim0.06-0.4$, the
luminosity would be orders of magnitude larger than that observed.

Narayan, Yi \& Mahadevan (1995) came up with an explanation for the
unusually low luminosity.  Building on previous work by Narayan \& Yi
(1994, 1995ab) and Abramowicz et al. (1995), they suggested that Sgr
A* may be accreting via an advection-dominated accretion flow (ADAF).
The key feature of an ADAF is that the gravitational energy that is
released as gas sinks in the potential well of the BH remains locked
up in the gas as thermal energy (or entropy) instead of being
radiated.  This can happen if the gas has a low enough density, as is
expected at low mass accretion rates $\dot M$, so that the charged
particles in the gas rarely scatter off one another and therefore do
not radiate very much.  The result is that the gas accretes down to
the center with a very high temperature ($\sim10^{12}$ K).  When the
gas reaches the BH horizon, it simply falls in with all its thermal
energy, and the accretion flow is extremely dim.

The above explanation for Sgr A*'s unusually low luminosity obviously
requires the source to have an event horizon.  If Sgr A* were to have
a surface, then the accreting hot gas would come to a stop when it
hits the surface and would radiate all its stored thermal energy
(since the density would go up by a large factor and the radiative
efficiency would increase correspondingly).  Therefore, to understand
the extraordinary dimness of Sgr A*, we require the object to be a
true BH with an event horizon (Narayan et al. 1998).  Note that Sgr A*
is not an isolated example.  The vast majority of supermassive BHs in
nearby galaxies are very underluminous relative to the gas supply
available to them (Fabian \& Canizares 1988).  The same explanation,
viz., an ADAF accreting onto an event horizon, will work for all these
objects (Fabian \& Rees 1995).

One major complication with this argument is that the gas in an ADAF
is very weakly bound to the BH because of its large temperature.  It
therefore tends to be ejected from the system rather easily (Narayan
\& Yi 1994, 1995a; Blandford \& Begelman 1999; Quataert \& Narayan
1999).  As a result, less gas reaches the BH than one might naively
expect given the amount of gas supply, and so the luminosity deficit
is less serious than one might have originally thought.
Quantitatively, of the luminosity deficit of $\sim10^{-4}$ seen in Sgr
A*, roughly $10^{-2}$ is from gas ejection and the remaining
$\sim10^{-2}$ is the result of advection and the disappearance of
energy through the event horizon.  The argument for the event horizon
thus survives, but it is weakened a little.

What one would really like to do is compare ADAFs around two accreting
objects, one with a surface and one with a suspected event horizon.
In such a comparison, if the object with a surface is much brighter
than the other object, then the case for an event horizon in the
latter would be very strong.  Narayan, Garcia \& McClintock (1997)
suggested that precisely such a comparison could be done using XRBs.

As mentioned in \S1, most of the known stellar-mass BH candidates are
in a class of XRBs called X-ray novae.  These systems are
characterized by a variable mass accretion rate, and tend to spend
most of their time in a quiescent state with a very low $\dot M$ and
$L_{\rm acc}$.  Only occasionally do they go into outburst, when they
accrete with high $\dot M$ and become bright.  Spectral observations
of quiescent BH XRBs can be explained in terms of an ADAF (see
Narayan, Mahadevan \& Quataert 1998 for a review).  Interestingly, in
addition to the many X-ray novae that have been identified with BH
candidates, there are other X-ray novae in which the accreting object
is a NS.  These latter systems too go into long periods of quiescence,
when they presumably have ADAFs.  Narayan et al. (1997) suggested that
by comparing the quiescent luminosities of BH and NS X-ray novae, one
could test for the presence of event horizons in the former: the BH
candidates should be much dimmer than the NSs.  The very limited data
available in 1997 seemed to bear this prediction out.

The observational situation has improved considerably since 1997 (see
Garcia et al. 2001; Hameury et al. 2003; McClintock et al. 2004), and
one can now state with considerable confidence that quiescent BH X-ray
novae are very much dimmer than quiescent NS X-ray novae.  The
luminosity difference is a factor of 1000 when measured in Eddington
units, and a factor of 100 when compared directly without Eddington
scaling.  Figure 3 shows the latest data on quiescent luminosities
plotted against the binary orbital period $P_{\rm orb}$.  The mass
accretion rate is determined largely by the latter quantity (Lasota \&
Hameury 1998; Menou et al. 1999) and hence NS X-ray novae and BH X-ray
novae with similar orbital periods are likely to have similar mass
accretion rates.  The fact that their luminosities are so different is
then both surprising and significant, and must reflect some
fundamental difference between the two kinds of object.  While a few
counter-explanations have been proposed (see Narayan, Garcia \&
McClintock 2002; Hameury et al. 2003; McClintock et al. 2004 for
references and rebuttals), the most natural explanation of the
observations is that accretion in quiescent XRBs occurs via an ADAF;
the BH candidates are dimmer because they have event horizons
through which they swallow all the advected energy.  In other words,
{\it black holes really are black} relative to other compact objects
with surfaces!

A related argument is based on the X-ray spectra of quiescent X-ray
novae.  Most quiescent NSs have X-ray spectra consisting of two
distinct components: a power-law component and a thermal
blackbody-like component.  The former is identified with a hot (likely
advection-dominated) accretion flow, and the latter with emission from
the surface of the NS.  McClintock et al. (2004) analyzed the spectrum
of the BH X-ray nova XTE J1118+480 in quiescence and established a
very tight limit on a thermal component: $L_{\rm th} <
9.4\times10^{30} ~{\rm erg\,s^{-1}}$ at the 99\% confidence level.
(There is a very clear power-law component in the spectrum, as one
would expect for a hot ADAF.)  In comparison, the thermal component in
NS systems is typically much brighter, $\sim {\rm few} \times 10^{32}$
to ${\rm few}\times10^{33} ~{\rm erg\,s^{-1}}$.  Since the thermal
component in NSs is identified with surface emission, the lack of a
similar component in XTE J1118+480 argues for the lack of a surface in
this source, i.e., the object must have an event horizon.  This
independent argument based on the spectrum bolsters the previous
discussion, which was based purely on luminosity.

\subsection{Absence of a Boundary Layer}

When an accretion disk, especially one with relatively cool gas in the
``high soft state,'' is present around a compact star, a narrow
viscous boundary layer will form at the radius where the rapidly
orbiting gas meets the surface of the star.  A considerable amount of
heat energy is expected to be liberated in this boundary layer.  No
boundary layer is expected if the central object is a black hole.

Sunyaev \& Revnivtsev (2000) showed that the variability power spectra
of NS XRBs have significant power even at frequencies as high as 1
kHz, whereas the power spectra of BH XRBs decline strongly above about
50 Hz.  Part of the difference is explained by the mass difference
between the two objects, since characteristic dynamical frequencies
scale as $M^{-1}$.  But even after allowing for this effect, there is
still a residual difference in the cutoff frequencies.  Sunyaev \&
Revnivtsev suggested that NSs have an additional high frequency
component in their variability spectra coming from the boundary layer
region, whereas the BH candidates lack this component.  If correct,
this argument implies that BH candidates must have event horizons.

In related work, Done \& Gierlinsky (2003) studied the X-ray spectra
of bright BH XRBs and NS XRBs.  They defined a ``hard color'' and a
``soft color'' to characterize the observed spectra and plotted these
quantitites for BH and NS systems in a color-color diagram.  They
showed that there are several similarities between BHs and NSs, but
also some clear differences.  In particular, a certain kind of pattern
in the color-color diagram is seen only from NS systems.  Done \&
Gierlinsky interpreted this unique spectral component as emission from
a boundary layer.  If their identification is correct, then the lack
of a similar component in BH XRBs implies that a boundary layer is
missing, i.e., these objects must have event horizons.

\subsection{Type I X-ray Bursts}

When gas accretes on the surface of a neutron star, it becomes denser
and hotter as it sinks under the weight of continued accretion, until
at a certain depth it ignites thermonuclear reactions.  The ignited
reactions are usually unstable, causing the accreted layer of gas to
burn explosively within a very short time.  After the fuel is
consumed, the star reverts to its accretion phase until the next
thermonuclear instability is triggered.  The star thus undergoes a
semi-regular series of thermonuclear explosions (see Lewin, van
Paradijs \& Taam 1993; Bildsten 1998; for reviews).  These
thermonuclear bursts, called Type I bursts, were first discovered from
X-ray binaries by Grindlay et al. (1976).  

In a typical Type I X-ray burst, the luminosity of the neutron star
increases to nearly the Eddington limit in less than a second, and the
flux then declines over a period of seconds to tens of seconds.  The
time interval between bursts is usually several hours to perhaps a day
or two.  The physics of Type I bursts has been widely studied, and the
broad features of the phenomenon are understood.  Theoretical models
generally agree quite well with observations (Lewin et al. 1993;
Bildsten 1998; Narayan \& Heyl 2003).

Remarkably, no Type I burst has ever been seen in any BH XRB (e.g.,
Tournear et al. 2003).  Why should this be the case?  Calculations
indicate that if a BH candidate with (say) a mass of $10 M_\odot$ were
to have a surface, it would exhibit Type I bursts with a burst rate
only a factor $\sim 3$ smaller than for a NS (Narayan \& Heyl 2002;
Yuan, Narayan \& Rees 2004).  Yet, although thousands of bursts have
been observed from NSs, not one burst has been seen from any BH
candidate!

Narayan \& Heyl (2002) argued that the lack of bursts is strong
evidence for the presence of horizons in BH candidates.  Clearly, if
the object has no surface, then gas cannot accumulate and cannot
develop a thermonuclear instability.  But is the lack of a surface the
{\it only} reason why BH candidates do not have bursts?  Is there any
reasonable scenario in which the objects might have surfaces and still
not burst?  There has been some discussion of this point (see Narayan
2003; Yuan et al. 2004; also Abramowicz, Kluzniak \& Lasota 2002).  In
brief, it appears that one needs to invoke very unusual physics if one
wishes to explain how BH candidates could have surfaces and yet not
produce Type I bursts.  In fact, the requirements are so extreme that
it is far more economical simply to accept that BH candidates have
event horizons!

\subsection{Direct Imaging}

Although none of the arguments described above is absolutely rigorous,
in combination they leave almost no room for any model of BH
candidates that does not include an event horizon.  Thus, it is fair
to say that there is a compelling case for the presence of event
horizons in astrophysical BHs.  The evidence is however indirect, and
one wonders whether it is possible to obtain more direct evidence.

The most promising idea is to obtain an image of the region near the
event horizon of an accreting BH.  At first sight this seems
impossible, considering how compact astrophysical BHs are and how far
away they are from us.  However, the situation is not entirely
hopeless.  Consider for concreteness a non-rotating BH with a horizon
at radius $R_S$.  Because of strong light-bending in the vicinity of
the BH, a distant observer will see an apparent boundary of the BH at
a radius of $(27)^{1/2}R_S/2$ (Falcke, Melia \& Agol 2000).  Rays with
impact parameters inside this boundary intersect the horizon, while
rays outside the boundary miss the horizon.  The angular size of the
boundary is
\begin{equation}
\theta_b = {(27)^{1/2}R_S\over 2D} = 5\times 10^{-8} \left({M\over
1M_\odot}\right) \left({1~{\rm kpc}\over D}\right) ~{\rm mas},
\end{equation}
where mas = milliarcsecond = $4.85\times10^{-9}$ radian.  A
$10M_\odot$ BH candidate at a distance of 1 kpc will have $\theta_b
\sim 10^{-6}$ mas, which is much too small to be resolved with any
technique in the foreseeable future.  However, Sgr A* with $M\sim
4\times10^{6}M_\odot$ and at a distance of 8 kpc should have an
angular size of $\sim 0.02$ mas which is not beyond reach.  The
supermassive BH in the nucleus of the nearby (15 Mpc) giant elliptical
galaxy M87, with a mass of $3\times10^9M_\odot$ and an expected
angular size of $\sim 0.01$ mas, is another object of interest.

The best angular resolution achievable today is with radio
interferometry, where angles less than 1 mas are routinely resolved.
In the not too distant future, it should be possible to operate
interferometers at wavelengths $\lambda < 1$ mm and with baselines as
large as the diameter of the earth $b\sim10^4$ km.  The nominal
angular resolution of such an interferometer would be $\lambda/b \sim
0.02$ mas, i.e., suitable for resolving the image of Sgr A*.
Interestingly, Sgr A* radiates most of its (very small) luminosity in
sub-millimeter waves, so there will be a measurable signal at these
wavelengths.  Moreover, there is a reasonable chance that the
accreting gas will be optically thin so that one will be able to see
through it.  In addition, since the gas is advection-dominated it is
likely to be quasi-spherical rather than disk-like, thus enabling us
to image all around the BH.

Falcke et al. (2000) suggested that by imaging Sgr A* with
sub-millimeter radio interferometry, one could check whether there is
a reduction in the surface brightness inside $\theta_b$.  They call
this the ``shadow of the black hole.''  The detection of this shadow
would certainly confirm a key prediction of general relativity.  But
will it constitute proof of the event horizon?  That is not obvious.
The shadow of the BH is caused by strong gravitational bending of
light rays; in particular, it reflects the existence of circular
photon orbits, which is an effect of strong gravity.  However, even a
NS, if it were sufficiently compact, would have photon orbits.
Therefore, if a shadow is seen in sub-millimeter waves in Sgr A*, all
that it means is that there is no significant radiation coming from
the surface of the BH at these wavelengths.  This could be because the
BH has no surface, i.e., it has an event horizon, or it could simply
be that the object has a surface that happens not to radiate in this
band.

Looking further into the future, there are ambitious plans to develop
X-ray interferometry in space (e.g., the MAXIM mission concept, Cash
2002) to study accretion flows around supermassive black holes in
other galactic nuclei.  If these plans come to fruition, one could
imagine mapping disks in say the X-ray iron line and watching the
image change with time (e.g., Movie3 and Movie4)!  Such observations
are probably at least a few decades away.

\section{Extracting Energy from Spinning Black Holes}

Astrophysical BHs are expected to have non-zero angular momentum.  The
progenitors of stellar-mass BHs almost certainly were spinning when
they collapsed, so their end-products should be spinning.
Supermassive BHs too should have considerable angular momentum since
they are believed to have acquired most of their mass from an orbiting
accretion disk.  A spinning BH has free energy associated with its
rotation which can in principle be tapped, e.g., by the Penrose
process.  Given all this, it has been suspected for long that some of
the most energetic phenomena associated with astrophysical BHs may
somehow be related to BH spin.  However, convincing evidence has been
hard to come by.

Model fits of the iron line in some sources require a very steep
emissivity function (see Reynolds \& Nowak 2003); most of the energy
has to be emitted closer to the BH than one expects if the energy
source is the accretion flow.  This has led to the suggestion that
perhaps a substantial fraction of the line energy is supplied by the
BH, although there is as yet no convincing model showing how this
happens.

Another promising possibility is that relativistic jets, which are so
ubiquitous in accreting BHs, are launched from the ergospheres of
rotating BHs.  Jets with very large bulk Lorentz factors
$\gamma\sim10$ have been known for long in radio-loud AGN (Zensus
1997).  More recently, moderately large Lorentz factors $\gamma\sim$
few have been seen in BH XRBs (Mirabel \& Rodr\'iguez 1999).  In
addition, there is clear evidence for jets with $\gamma\sim100$ in the
ejecta of gamma-ray bursts (GRBs, M\'esz\'aros 2002).  GRBs are known
to be associated with the deaths of massive stars in ``hypernovae,''
and while there is no certainty that these explosions produce BHs at
their centers, there is a good chance that they do.

The original idea of Penrose for extracting energy from a rotating BH
involves a particle that enters the ergosphere and then splits into
two.  One of the two particles falls into the BH on a negative-energy
orbit, causing the mass, energy and angular momentum of the BH to
decrease, while the other particle escapes to infinity carrying the
excess energy and angular momentum with it.  This process requires
particular conditions (Bardeen, Press \& Teukolsky 1972) that are
probably not common in nature.  Another process to extract energy from
a rotating BH is via ``super-radiance'' in which a a suitably tuned
wave impinges on the BH and is reflected with increased energy (Press
\& Teukolsky 1972).  This mechanism has been suggested as an energy
source for GRBs (van Putten 1999; Aguirre 2000).  Presently, the most
promising idea for extracting energy from a rotating BH is via
magnetic fields (Znajek 1976; Blandford \& Znajek 1977) since fairly
strong magnetic fields are likely to be present in all accretion
flows.  Magnetic fields are capable of connecting regions very close
to the BH to regions farther out, thereby introducing considerable
long-range coherence.


How exactly does the magnetic field tap the energy of a rotating BH?
Recent general relativistic numerical MHD simulations are beginning to
provide an answer.  Koide et al. (2002) and Koide (2004) describe
simulations of a magnetized plasma in the ergosphere of a spinning BH.
They find that, near the equatorial plane, the field lines are
azimuthally twisted because of the dragging of intertial frames by the
spinning hole.  The twist then propagates outward and transports
energy along twin jets that form naturally.  The results are in
general agreement with an earlier proposal made by Punsly \& Coroniti
(1990).  In addition, Koide et al. (2002) have shown that the energy
extraction mechanism is related to the Penrose process.  As the field
lines in the equatorial region are pulled forward, the local plasma on
the field line acquires a negative energy.  This plasma falls into the
BH, while a torsional Alfven wave is driven outward along the field
lines, generating an outward Poynting flux and at the same time
increasing the energy of plasma outside the ergosphere.  The authors
call it the {\it MHD Penrose process} since there is a close analogy
between this mechanism and Penrose's original idea.  The main
difference is that the negative energy particles close to the BH and
the positive energy particles farther out communicate via magnetic
field lines, which act as energy conduits.

Semenov, Dyadechkin \& Punsly (2004) have recently used a simplified
set of equations to follow the evolution of individual magnetic flux
tubes in the vicinity of a rotating BH.  Movie5 and Movie6 from their
work show how a rotating BH with a magnetized plasma around it
naturally develops a configuration with infalling negative energy
plasma in the equatorial plane and outgoing coiled magnetic fields
that carry away energy in twin jets.  Much more work is obviously
needed, but it appears that the connection between rotating BHs and
jets is beginning to be understood.

Note that the field topology in the above-cited work is different from
the traditional picture in which field lines connect the BH horizon to
the accretion disk.  The traditional field configuration seems to
produce evacuated funnels with outflowing funnel walls rather than
bona fide jets (de Villiers, Hawley \& Krolik 2003; McKinney \& Gammie
2004).  The simulations of Koide et al. (2002) and Semenov et
al. (2004) start out with field lines going from the ergosphere to
infinity along the spin axis.  Such a field configuration can be
achieved naturally as a result of magnetized plasma falling into the
BH (Narayan, Igumenshchev \& Abramowicz 2003).  In fact, not only does
this field configuration enable energy extraction from a rotating
hole, it may also enhance the energy extracted from gas accreting onto
a {\it non-rotating} hole, i.e., it can give a larger value of $\eta$
than the standard value of 0.057 shown in Fig. 2 (Narayan et
al. 2003).

\section{Conclusion}

Astrophysicists are contributing in important ways to the study of
BHs.  They are discovering increasing numbers of BH candidates, they
are measuring the fundamental parameters of the BHs, they are finding
strong evidence that BH candidates have event horizons, and they are
beginning to carry out additional tests of relativity in the regime of
strong gravity.  However, all the work done so far relates only to the
classical regime.  Quantum effects have remained out of reach.

The most famous quantum effect is, of course, Hawking (1974)
radiation.  Unfortunately, for massive BHs such as the ones discovered
so far, the Hawking temperature is so small [$\sim 10^{-7}(M_\odot/M)$
K] that there is no hope of ever detecting this radiation.  The story
is different in the case of BHs with $M<10^{15}$ g.  Such BHs will
evaporate by the Hawking mechanism in a time less than the age of the
universe and will explode with final luminosities $\sim 10^{20} ~{\rm
erg\,s^{-1}}$ (Shapiro \& Teukolsky 1983).  If the universe made such
mini-BHs in sufficient numbers, we might expect to see a few of them
die in fleeting bursts of high energy emission.  Given our present
understanding of the formation of compact objects by stellar collapse,
and given the constraints on density fluctuations in the early
universe as determined from observations of the cosmic microwave
background, there does not seem to be much chance of finding
$10^{15}$g BHs in the universe today.  However, nature has a long
history of surprising us --- nowhere more true than in astrophysics
--- so perhaps we should not give up hope altogether.

\acknowledgments

This work was supported in part by NASA grant NAG5-10780 and NSF grant
AST 0307433.

\bibliography{}

\newpage
\noindent
{\bf Movie Captions}

\bigskip\noindent Movie1.--- Shows the orbits of individual stars near
the Galactic Center as measured with high resolution infrared
observations.  The movie runs from 1992, the initial date of the
observations, to the present, and is extrapolated a few years into the
future.  Time is shown at top left.  The star whose track (with error
bars) traces a complete ellipse was fitted by Sch\"odel et al. (2002)
to a highly elliptical Keplerian orbit.  From the fit they calculated
the mass of the supermassive BH to be $3.7 \pm 1.5 \times 10^6
M_\odot$.  The inferred position of the BH is shown by the red cross.
(Movie courtesy Reinhard Genzel)

\bigskip\noindent Movie2.--- Shows the orbits of individual stars near
the Galactic Center as measured with high resolution infrared
observations.  The movie runs from 1995, the initial date of the
observations, to the present, and is extrapolated a few years into the
future.  Time is shown at top left.  By combining the stellar
positions with Doppler radial velocity measurements and fitting to
Keplerian orbits, Ghez et al. (2004) refined the mass estimate of the
supermassive BH to $3.7 \pm 0.2 \times 10^6 M_\odot$.  The inferred
position of the BH is shown by the stationary * at the center.  (Movie
courtesy Andrea Ghez)

\bigskip\noindent Movie3.--- Shows the simulated image of a turbulent
accretion disk heated by the dissipation of magnetic fields.  The dark
region at the center is the inner boundary of the simulation, which is
at a radius of $2R_S$.  During the sequence the view changes from
almost face-on ($i=1\degr$) to nearly edge-on ($i=80\degr$), as
indicated by the bar on the right ($0\degr$ is at the bottom of the
bar and $90\degr$ at the top).  As the inclination increases, note how
the emission becomes enhanced to the left of the BH because of Doppler
boost.  Also, even though the disk is perfectly flat, it appears to be
warped upward behind the BH.  This is because of the deflection of
light rays by the gravity of the BH.  (Based on Armitage \& Reynolds
2003; movie courtesy the authors)

\bigskip\noindent Movie4.--- Shows the variation of the fluorescent
iron line in the simulation shown in Movie3.  The inclination is held
fixed at $80\degr$ and the line emissivity is taken to be proportional
to the local energy generation rate.  Note that the line profile,
shown at bottom right, varies rapidly as a result of turbulent
fluctuations in the disk.  The line extends from about 4 keV on the
left to about 7 keV on the right, with a peak at around 6 keV.  (Based
on Armitage \& Reynolds 2003; movie courtesy the authors)

\bigskip\noindent Movie5.--- Simulation of a magnetic flux tube
accreting onto a maximally rotating BH (Semenov et al. 2004).  The
light circle at the center represents the event horizon of the BH, and
the shaded region around it is the ergosphere.  As the field line is
drawn in by the gravity of the BH, it is pulled forward azimuthally by
the dragging effect of the BH spin.  As a result, some of the plasma
near the equatorial plane, shown in red, acquires negative energy as
viewed from infinity.  When this gas falls into the BH, it effectively
reduces the energy and angular momentum of the BH.  Correspondingly,
electromagnetic and plasma energy is ejected along twin jets that move
out parallel to the spin axis of the BH.  Koide et al. (2002) named
this the MHD Penrose process of extracting energy from a rotating BH.
(Movie from Semenov et al. 2004, courtesy Brian Punsly)

\bigskip\noindent Movie6.--- Expanded view of the simulation shown in
Movie5.  Note the dramatic coiled magnetic field in the two outgoing
jets.  The jets are powered by the spinning BH via the MHD Penrose
process.  (Movie from Semenov et al. 2004, courtesy Brian Punsly)

\newpage

\noindent
{\bf Figure Captions}

\bigskip\noindent Fig. 1.--- [Upper panel] The red, blue and green
spots show the relative locations of maser emitting clouds in the gas
disk at the center of NGC 4258, as determined with radio
interferometry.  The black dot is the calculated position of the
central supermassive BH.  Emanating from the BH are the beginnings of
twin jets.  The dashed lines trace out a model of the disk.  [Bottom
panel] The dots show the line-of-sight velocities of the maser spots
as a function of distance from the BH along the major axis of the
disk.  Note how perfectly the velocities follow the Keplerian profile
traced by the line.  By fitting the velocities, the mass of the BH is
inferred to be $3.5\pm0.1\times 10^7M_\odot$.  (Based on Greenhill et
al. 1995; Herrnstein et al. 1998; image courtesy Lincoln Greenhill)

\newpage

\begin{figure}
\figurenum{2}
\plotone{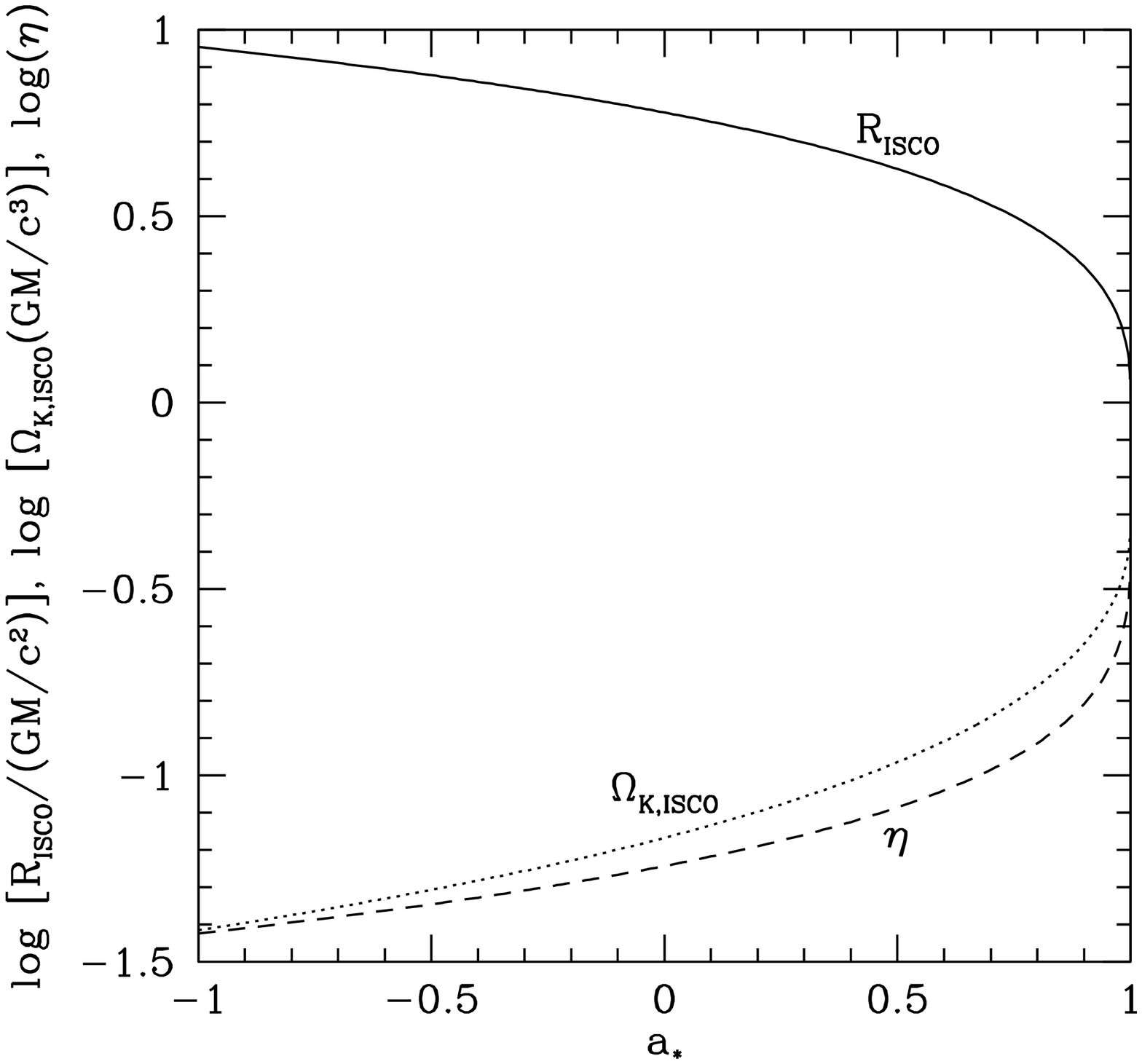}
\caption{The radius of the innermost stable circular orbit $R_{\rm
ISCO}$, the Keplerian frequency at this radius $\Omega_{\rm K,ISCO}$,
and the binding energy at this radius $\eta$, as functions of the BH
spin parameter $a_*$.  Positive values of $a_*$ imply that the BH
corotates with the orbit and negative values mean that the BH
counter-rotates.  By measuring the quantity $R_{\rm ISCO}/M$ or
$\Omega_{\rm K,ISCO}M$ or $\eta$, one could estimate $a_*$.}
\end{figure}

\newpage
\begin{figure}
\figurenum{3}
\plotone{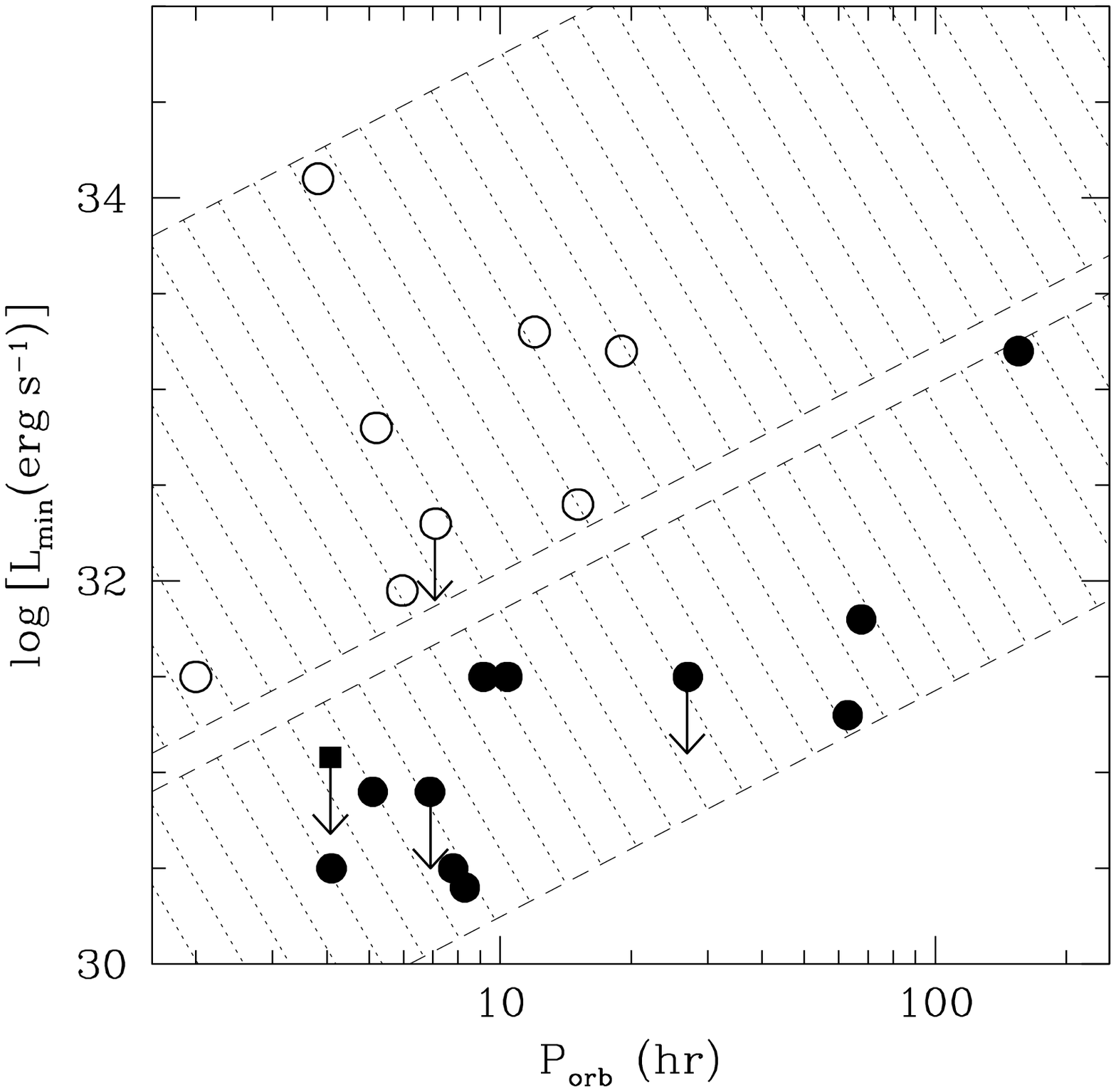}
\caption{Quiescent luminosities $L_{\rm min}$ of X-ray novae plotted
against the orbital period $P_{\rm orb}$ of the binaries.  The filled
symbols correspond to BH candidates, the open symbols to NSs, and
arrows represent upper limits.  The shaded bands are to guide the eye.
At any given orbital period, the NSs as a group are a factor of
$\sim100$ brighter than the BH candidates.  This difference may be
interpreted as evidence that the BH candidates possess event horizons.
(Taken from McClintock et al. 2004)}
\end{figure}

\end{document}